\documentclass[dvips]{article}
\usepackage{setspace}
\usepackage{amsmath,amssymb,amsthm,graphicx,latexsym}
\begin{document}
\begin{center}
\Large{\bf {"Stringy"  Coherent States Inspired By Generalized Uncertainty
Principle  }}\\
\vskip 1cm
Subir Ghosh and Pinaki Roy\\
\vskip .5cm
Physics and Applied Mathematics Unit\\
 Indian Statistical Institute\\
 203 B. T. Road, Kolkata 700108, India
\end{center}
\vskip .3cm
{\bf {Abstract:}} 
  Coherent
States with Fractional Revival property, that explicitly satisfy the Generalized
Uncertainty Principle (GUP), have been constructed in the context of Generalized
Harmonic Oscillator. The existence of such states is essential in motivating the
GUP based
phenomenological results present in the literature which otherwise would be of
purety academic interest. The effective phase space is Non-Canonical (or
Non-Commutative in 
popular terminology). Our results have a smooth  commutative
limit, equivalent to Heisenberg Uncertainty Principle. The
Fractional Revival time analysis yields an independent bound on the GUP
parameter.
Using this and similar bounds obtained here, we derive the largest possible
value of the (GUP induced) minimum
length scale. Mandel parameter analysis shows that the statistics is
Sub-Poissionian. Correspondence
Principle is deformed in an interesting way. Our computational scheme is very
simple as it requires only
first order corrected energy values and undeformed basis states.
\vskip 1cm
\section{Introduction:}
In recent years Generalized Uncertainty Principle (GUP) \cite{gup,gup1} has
created a
lot of excitement because  the works \cite{ncr1,ncr} have established that
Quantum Gravity signatures are quite
universal and GUP can serve as a (moderate energy) window to the latter. 
Contrary to
previously held belief, it can lead to small but observable effects at energy
scales considerably smaller than Planck energy. The GUP is strikingly distinct
from  Heisenberg Uncertainty Principle (HUP) since GUP unambiguously
{\it{fixes the smallest size of the dispersion}} $(\Delta
x)={\sqrt{<x^2>-<x>^2}}$. On the other hand in HUP either of $\Delta x$ or
$\Delta p$ can
arbitrarily small at the expense of the other. In a
way GUP points to a string like elementary excitations instead of
point particle. Furthermore GUP is compatible with a modified Non-Commutative
(NC) \cite{nc,nc1} symplectic structure that can have non-trivial effects in all
areas of
quantum physics.

{\it{But the most important player - an actual quantum state obeying GUP - is
missing
 in all the above analysis }} \cite{gup,ncr1,ncr}. These works  naively assume
the GUP itself and do not address, let alone show, that  quantum states obeying
GUP  and stable under time-evolution 
can exist. Compare this situation with
the conventional case where one can construct Coherent States (CS) for Harmonic
Oscillator (HO) that are stable and does saturate the HUP. In the present Letter
we have provided this vital missing information. In the  framework of
Generalized CS (GCS) \cite{gcs} we have
constructed the first order (in $\beta $- the GUP parameter) corrected GCS  that
satisfies GUP  and shows the essential {\it{Fractional Revival}} behaviour
indicating time-stability. For
$\beta =0$ the conventional CS for HO is recovered. Another major result of our
work is that we have provided an {\it{independent bound on}} $\beta $ that
agrees with \cite{ncr1}. Correspondence
Principle is maintained albeit in an interestingly modified way.  Lastly we have
developed a simple algorithm to show how first order corrected energy values can
yield 
first order corrected wave functions (GCS) endowed with non-trivial features of
the perturbation (GUP and NC physics, in the present case). Indeed, the
flexibility in GCS
construction \cite{gcs,kl} allows this act which can be generalized in many
interesting systems with non-linear
(compared to HO) energy dispersion.

GUP is probably unavoidable at high energy scales where Quantum Gravity prevails
since it 
supplies an energy scale $\sim$ Planck energy $\sim$  in an   inherent way.
GUP was predicted \cite{gup1} quite long ago from String
Theory
considerations and the inherent length scale in GUP intuitively resolves
\cite{nc} the
paradoxes in Black Hole physics. But GUP is directly linked
with a Non-Canonical (popularly referred to as Non-Commutative (NC)),  phase
space symplectic structure \cite{nc,nc1}. NC phase space has pervaded
theoretical physics
in recent years since a non-trivial change in the fundamental sysplectic
structure can affect physics
in major ways. NC phase space appears both in relativistic and non-relativistic
scenarios and the one relevant
to GUP belongs to the latter but can be reduced from the former \cite{nc1}.

GCS    are best suited for our purpose since one can quickly grasp the
non-classical behaviour of the system being studied. Ever since its introduction
by Schr\"odinger in the context of Harmonic
Oscillator
(HO), CS have played an ever increasing role in diverse
areas of physics. Ideal lasers enjoying Poissonian photon number statistics are
described by
Glauber-Sudarshan CS but to deal with non-linear interactions in a real laser
and
 furthermore in more
exotic theoretical frameworks (such as Quantum Group ideas \cite{qg} to describe
a real laser \cite{kat}) various
 generalizations of the former canonical CS, compatible with non-Poissonian
statistics, have been introduced. We shall follow one particular form of
Generalized
Coherent state (GCS), the so called Gazeau-Klauder CS, proposed in
\cite{kl,gcs} whose central idea is to construct GCS
that are stable in time (i.e GCS will remain GCS throughout its evolution). This
property is very
relevant in experimental context since this can explain the experimentally
observed fascinating phenomenon of fractional revival: an
initially well localized wave packet disperses due to quantum effects but after
sufficient time, depending
on the parameters of the problem, recombines to form the initial state once
again (revival) but during the process the state becomes multi-localized thus
 exhibiting fractional revival. The latter state appears as a collection of
sub-wave packets separated in space but
all resembling the initial wave packet. These fractional revival have shorter
time scales than the revival time. The fractional
revivals are a manifestation of the non-linearity in the energy spectrum
(compared to the linear spectrum
$E_n \sim n $ in HO) which is precisely the reason of introducing GCS that
maintain time stability. In NC physics
context CS have appeared in 
\cite{nic} in NC extended Black Hole  and GCS in \cite{sch} in NC Moyal plane,
with which
we will compare our results.

In this Letter we consider a Generalized HO (GHO) whose coordinate and momentum
satisfy a
Non-Commutative (NC) algebra
that induces the GUP. The non-commutative (or GUP) parameter $\beta $ brings in
the minimum length scale. At the same time $\beta $ changes the HO
spectrum to a non-linear one
thus ushering in the GCS. We show that the fractional revival occurs only for
non-zero $\beta $ and
hence its experimental (non-) observation can clearly put bounds on $\beta $.\\
\section{ NC phase space, GUP and GHO}
 We consider the following
Non-Commutative (Jacobi
identity satisfying)
two-dimensional phase space, leading to the GUP,
\begin{equation}
 [x,p]=i\hbar (1+\beta p^2)~; [x,x]=[p,p]=0~; 
\rightarrow \Delta x \Delta p \geq \frac{\hbar}{2}(1+\beta A(\Delta p)^2).
\label{2}
\end{equation}
where $\beta =l_{PL}^2/(2\hbar ^2),~l_{PL}^2=G\hbar /c^3\equiv (Planck
~length)^2$. 
Throughout the Letter we will keep only $O(\beta )$ corrections. $A$ is a
$\beta$-independent
numerical constant. The RHS can be of a slightly more general form.  Let us
exploit the Darboux
map \cite{ncr1} (for the exact relativistic map see Ghosh and Pal in \cite{nc1},
\begin{equation}
 x\equiv X~,~p=P(1+\beta P^2),
\label{01}
\end{equation}
where $X,P$ are canonical, $[X,P]=i\hbar,~[X,X]=[P,P]=0$. This is different from
the conventional map where one deforms $x$.

The GHO
\begin{equation}
 H=\frac{p^2}{2m} + \frac{m\omega^2x^2}{2}
\label{02}
\end{equation}
in the canonical $X,P$ coordinates becomes
\begin{equation}
 H=\frac{P^2}{2m}(1+\beta
P^2)^2+\frac{m\omega^2X^2}{2}\approx
\frac{P^2}{2m}+\frac{m\omega^2X^2}{2}+\frac{\beta
P^4}{m}.
\label{03}
\end{equation}
In \cite{ncr1} this $H$ was used to compute $O(\beta )$ corrections for
different potentials. We introduce canonical creation-annihilation
operators:
\begin{equation}
 a=\sqrt{\frac{m\omega}{2\hbar}}(X+i\frac{P}{m\omega})~,~a^\dagger
=\sqrt{\frac{m\omega
}{2\hbar}}(X-i\frac{P}{m\omega}) \rightarrow [a,a^\dagger]=1;
[a,a]=[a^\dagger,a^\dagger]=0.
\label{04}
\end{equation}
Our scheme is such that, in our level of approximation, it is enough to
use the canonical HO
Fock basis:
\begin{equation}
a\mid 0>=0~;~~a\mid n>=\sqrt{n}\mid n-1>~,~~a^\dagger \mid n>=\sqrt{n+1}\mid
n+1>.
\label{fock}
\end{equation}
The Hamiltonian turns out to be,
\begin{equation}
 H=\hbar\omega (a^\dagger a+\frac{1}{2})+\frac{\beta
m\omega^2\hbar^2}{4}(a-a^\dagger)^4.
\label{05}
\end{equation}
Terms with same number of $a$ and $a^\dagger$ will contribute to the energy
spectrum,
\begin{equation}
 E_n =\hbar\omega~[n(1+\lambda (1+n))+\frac{1}{2}(1+\frac{\lambda}{2})];~\lambda
\equiv (3\beta m\hbar \omega)/2.
\label{06}
\end{equation}
We drop the unimportant constant part in $E_n$ and use the notation of
\cite{gcs},
\begin{equation}
\hbar\omega e_n \equiv \hbar\omega n(1+\lambda(1+n) ).
\label{en}
\end{equation}
\subsection{Generalized Coherent States for GHO} This section constitutes the
main
body of our work. The Gazeau-Klauder  \cite{gcs} GCS are defined as
\begin{equation}
\mid J,\gamma >=\frac{1}{N(J)}\sum _{n\geq 0}\frac{J^{n/2}e^{-i\gamma
e_n}}{\sqrt
{\rho_n}}\mid n>~;~~\rho_0=1~,~~\rho_n=e_1e_2...e_n .
\label{gcs}
\end{equation}
where $J$ and $\gamma$ are two real numbers parameterizing the coherent state
along with 
the normalization condition $
N(J)^2=\sum _{n\geq 0}\frac{ J^n}{\rho_n}$ that can be expressed in terms of
modified Bessel Function (see {{\it {e.g.}} \cite{gcs}). 

Let us compute $x$ and $p$ dispersion  for the GCS $\mid J,\gamma >$ (note that
the physical variables are $x,p$ and not
the canonical $X,P$):
\begin{equation}
(\Delta x)^2 = <x^2>-(<x>)^2~; ~~(\Delta p)^2 = <p^2>-(<p>)^2.
\label{gup}
\end{equation}
A straightforward calculation reveals (see  Appendix for a brief outline of the
computation)
\begin{equation}
<x>= {\sqrt{\frac{\hbar }{2m\omega }}}<a+a^\dagger > $$$$
= {\sqrt{\frac{2\hbar J}{m\omega }}}[cos \gamma -\lambda
\{(1+\frac{J}{2})cos\gamma +2(1+J)\gamma sin\gamma \}],
\label{j1}
\end{equation} and in a similar way,
\begin{equation}
<x^2>=\frac{\hbar }{2m\omega }<a^2+a^{\dagger 2}+aa^\dagger +a^\dagger a > $$$$
=\frac{\hbar }{2m\omega }[1+2J+2Jcos(2\gamma ) -2\lambda J
\{2+J+(\frac{5}{2}+J))cos(2\gamma)+(6+4J)\gamma sin(2\gamma)\}]
\label{jj3}.
\end{equation} This leads to the $x$-dispersion,
\begin{equation}
(\Delta x)^2=\frac{\hbar}{2m\omega}[1-\lambda J \{cos (2\gamma )+4\gamma sin
(2\gamma)\}].
\label{dis}
\end{equation}
We compute analogous expressions for $p$,
\begin{equation}
<p> =-i{\sqrt{\frac{m\omega \hbar}{2}}}<a-a^\dagger
-\frac{\lambda}{3}(a-a^\dagger )^3> $$$$
={\sqrt {2m\omega \hbar J }}[-sin\gamma 
+\lambda \{-\frac{J}{2}sin\gamma +\frac{J}{3}sin(3\gamma )-2 (1+J)\gamma
cos\gamma \}],
\label{j3}
\end{equation}
\begin{equation}
 <p^2>=-\frac{m\omega
\hbar}{2}<(a-a^\dagger)^2-\frac{2\lambda}{3}(a-a^\dagger)^4> $$$$
=\frac{m\omega \hbar}{2}[1+2J-2Jcos(2\gamma )+\lambda
\{2+4J+2J^2+(3J-\frac{10}{3}J^2)cos(2\gamma )+
\frac{4}{3}J^2 cos(4\gamma )+2J(6+4J)\gamma sin(2\gamma )\}],
\label{0j3}
\end{equation}
and subsequently obtain the $p$-dispersion,
\begin{equation}
(\Delta
p)^2=\frac{m\omega\hbar}{2}[1+\lambda \{2+4J-3Jcos(2\gamma )+4J\gamma
sin(2\gamma )\}.
\label{disp}
\end{equation}
Hence the  product of the dispersions is,
\begin{equation}
(\Delta x)^2(\Delta p)^2=\frac{\hbar^2}{4}[1+2\lambda \{1+2J-2Jcos(2\gamma )\}].
\label{uncert}
\end{equation}
The energy of the GCS turns out to be,
\begin{equation}
<E>=\frac{1}{2m}(<p^2>+m^2\omega ^2<x^2>)=\frac{\hbar
\omega}{2}[1+2J+\lambda \{1-4J(1+\frac{2}{3}J)cos(2\gamma
)-\frac{2}{3}J^2cos(4\gamma )\}]
\label{euncert}
\end{equation}
Let us now compute the dispersion of the number operator $N=a^\dagger a$ :
\begin{equation}
<N> =<a^\dagger a>=J[1-\lambda (2+J)],~<N^2>= J(1+J)(1-2\lambda (1+J))
\rightarrow (\Delta N)^2=J(1-2\lambda ).
\label{n}
\end{equation}
\subsection{Fractional Revival} This section deals with the essential stability
criteria
of our GUP satisfying GCS. To discuss revival structure of the coherent states,
let us re-write the energy
eigenvalues as
\begin{equation}
E_n=an^2+bn,~~~~a=\lambda\hbar\omega ,~~~~~b=\hbar\omega(1+\lambda)\label{en}.
\end{equation}
Then the coherent state at a time $t$ is given by
\begin{equation}
|J,\gamma,t>=\frac{1}{N(J)}\sum_{n=0}^\infty \frac{J^{n/2}e^{-i(\gamma
e_n+E_nt/\hbar)}}{\sqrt{\rho_n}}|n>
\end{equation}
Next we consider the autocorrelation function defined by
\begin{equation}
A(t)=<J,\gamma,t|J,\gamma,0>
\end{equation}
Thus squared modulus of the autocorrelation function $|A(t)|^2$ gives the
probability of how far the coherent state resembles its original form. So, if at
time $t=t_R$, the coherent state regains its form at time $t=0$, then
$|A(t_R)|^2=1$ and $t_R$ is called the period of full revival. Now from the
relations
(\ref{en}) it follows that time of full revival is given by
\begin{equation}
t_R=\frac{2\pi\hbar}{a}
\label{tr}
\end{equation}
provided the following relation is satisfied
\begin{equation}
\beta m\hbar\omega=\frac{1}{r-1},~~~~r=integer\geq 2,
\label{bound}
\end{equation}
where we assume $\beta $ to be positive.
From the above relation we find that for the coherent state to exhibit full
revival, the frequency has to be chosen suitably. Furthermore, it
is known that for suitably fine tuned coupling the coherent states exhibit
fractional revivals
with periods $t_{fr} =\frac{q}{r}t_R$ where $q/r$
is a non reducible fraction of integers. In fact, from Fig 1 it can be
seen that there are quarterly revivals at $t=t_R/4,3t_R/4$. It may be of
interest to note that these fractional revivals do not exist in the
$\beta\rightarrow 0$ limit and are essentially a consequence of the presence of
a minimal length. 
\section{Discussion}  Below we present our results and their implications.\\
{\bf{1. GUP and Dynamics:}} First we explicitly show that the
GCS 
satisfy GUP. From (\ref{uncert}) we recover the generalized uncertainty
principle (with  the original
parameter $\beta$ restored):
\begin{equation}
 (\Delta x)(\Delta p)=\frac{\hbar}{2}[1+3\beta
(\frac{m\hbar\omega}{2})\{1+2J-2Jcos(2\gamma)\}].
\label{gup0}
\end{equation}

Notice that, quite remarkably, the $\beta$-correction term in r.h.s. is
identical to the $\lambda$-independent 
part of $<p^2>$ as obtained previously in (\ref{0j3}), $
<p^2>=\frac{m\hbar\omega}{2}[(1+2J-2Jcos(2\gamma))+O(\lambda )]$. Hence to
$O(\lambda )$ we
have shown that
\begin{equation}
 (\Delta x)(\Delta p)=\frac{\hbar}{2}(1+3\beta <p^2>).
\label{gup01}
\end{equation}
Indeed this is not completely the GUP (\ref{2}) since it has $(\Delta p)^2$ on
the r.h.s.
But from (\ref{j3}) and (\ref{0j3}) we might
argue that for the $\lambda$-independent terms. $\sqrt {<p^2>}~>~ <p> $. Thus at
very high momentum regime, where 
the correction terms become significant $<p^2>~ >>~ (<p>)^2~~\rightarrow (\Delta
p)^2\approx <p^2>$ and we roughly recover the GUP (\ref{2}) from (\ref{gup01}).
However, since from (\ref{disp}) $(\Delta
p)^2=\frac{m\omega\hbar}{2}(1+O(\lambda ))$ , we can also interpret
(\ref{gup01}) as 
\begin{equation}
 (\Delta x)(\Delta p)=\frac{\hbar}{2}(1+\beta_{eff} (\Delta p)^2),
\label{gup02}
\end{equation}
where $\beta_{eff}=3(1+2J)\beta$ with the oscillatory part averaged out. In this
respect $\beta_{eff}$ loses partly
its unversality status since it becomes $J$-dependent but $J$ being positive
$\beta_{eff}\geq\beta$. Since $J$ is related to the energy of the coherent
state, this identification  will eventually mean that particles having different
energies will observe slightly
different minimum length scale with the lower bound being determined by $\beta
$. However, this observation may not appear too
unexpected if we recall that in 
the Doubly Special Relativity scenario \cite{nc1}, (which induces the
Non-Commutative phase space and GUP (\ref{2})  of the
present form in 
certain limiting condition), the generalized spacetime
transformations become explicitly dependent on the energy and momentum of the
particle whose coordinates are being
transformed and so it is never possible to isolate the kinematics (spacetime
transformations) from the dynamical
content (particle energy and momentum). It seems that similar type of behaviour
reappears for a realistic coherent state obeying the GUP that we have
constructed.

The Heisenberg equation of motion  $\dot
{<A>}=\frac{i}{\hbar}<[H,A]>$ 
yields,
 \begin{equation}
\dot{<x>}={\sqrt{\frac{2\hbar\omega J}{m}}}[-sin \gamma 
+\lambda ((1+\frac{J}{2})sin \gamma -2\gamma cos \gamma -\frac{4J}{3}sin
(3\gamma ))].
\label{j2}
\end{equation}
Comparison with (\ref{j3}) shows a mismatch in numerical factors. For $\lambda
=0$ the exact
Correspondence Principle is recovered. Again below
are the Newton's equations in 
operator form with numerical mismatch:
\begin{equation}
\dot{<p>}={\sqrt{\frac{\hbar m\omega ^3}{2}}}<[a^\dagger a,(a-a^\dagger )
-\frac{\lambda}{3}[a^\dagger a,(a-a^\dagger )^3]>,
\label{j5}
\end{equation}
\begin{equation}
\ddot{<x>}={\sqrt{\frac{\hbar \omega ^3}{2m}}}<[a^\dagger
a,(a-a^\dagger )-
\frac{4\lambda}{3}[a^\dagger a,(a-a^\dagger )^3]>.
\label{j4}
\end{equation}
Finally we provide dynamics of the GCS,
\begin{equation}
\ddot{<x>}=-\omega ^2 (1-\frac{8\lambda }{3})<x> .
\label{jj4}
\end{equation}
 Hence the centre of the wave packet does  follow a HO  path {\it{but with an
effective reduced frequency}}. As we have advertized earlier, {\it{
Corespondence Principle is satisfied but with
a twist}}. It is
curious to note the charged planar HO in a perpendicular constant magnetic field
also undergoes a reduction 
in the effective frequency. Furthermore the condition  $(1-\frac{8\lambda
}{3})\geq 0$ 
produces same order of magnitude bound on the GUP parameter $\beta_0$ as
discussed below.\\
{\bf{2.  Mimimum Length Scale:}} 
In \cite{ncr1} the authors consider Landau levels in a GUP corrected scenario
and comparing with
(Scanning Tunnelling Microscope ) experimental data come up with a bound $\beta
_0
\leq 10^{50}$ for
the dimensionless parameter $\beta _0=
(M_{Planc}c)^2\beta $ and  $\omega_c \sim 10^3~
GHz$ ($\omega_c $ being the cyclotron
frequency of an electron in $10~T$ magnetic field).
This is relevant in our context as well since we also use
undeformed HO states in
constructing the GCS. But this can have interesting consequences as we now show.
The GUP (\ref{2}) yields the lower bound for $\Delta x \geq \hbar {\sqrt{A\beta
}}$ (the minimum length scale)
 that in the present case reduces to
\begin{equation}
 \Delta x \geq \hbar
{\sqrt{3(1+2J)\beta }}=\frac{\hbar
{\sqrt{3(1+2J)\beta_0
}}}{M_{Planck}c}.
\label{delx}
\end{equation}
Taking the largest allowed $\beta =10^{50}$ \cite{ncr1} and for an electron-mass
particle in our
HO with $\omega =\omega _c =10^3GHz$ we derive  the largest allowed value for
the minimum length scale
to be $\approx 10^{-9}~meter$. This value can be considerably smaller depending
upon $\beta $.  It is conceivable to suggest that these states
might be experimentally observed.\\
{\bf{3.  Non-Poission Statistics:}} With the help of (\ref{n}) we determine
the Mandel parameter
$Q=(\Delta n)^2/<n> -1$ that is a measure
of the deviation from $Q=0$,  the Poissionian distribution. $Q\ge 0 $ ($Q\leq 0
$) are termed
 as Super-Poissionian  (Sub-Poissionian) statistics. In the present case
\begin{equation}
 Q= -\lambda (4+J)
\label{q}
\end{equation}
{\it{showing that the statistics is Sub-Poissionian}}. This can be compared with
the
constant NC space GCS of \cite{sch} that
enjoys Poissionian statistics.\\
{\bf{4. Revival Time Induced Bound:}}  Rigorous estimates of the upper 
bound on $\beta_0<10^{36} $, taking quantum field theoretic (Lamb shift) measurements in to account, have
already been provided in \cite{ncr1}. In this section our aim is modest since we want only to 
ensure that the properties of the GCS constructed here can give certain bounds on $\beta_0$ which are 
compatible with \cite{ncr1}, albeit much weaker.

From (\ref{bound}) we can
obtain an upper bound on $\beta _0$
that is independent of the bounds existing in the literature. In this case
taking the minimum value of
$r=2$ yields $\beta _0\leq \frac{(M_{Planck }c)^2}{m\omega \hbar}\approx
10^{53}$. This is
comparable to the one derived in \cite{ncr1} that we have used before and
provides an independent consistency
check of the entire framework. On the other hand, exploiting the (\ref{tr}) one
can get a relation
\begin{equation}
t_R \beta _0\approx 10^{41},
\label{tr1}
\end{equation}
for the same parameters $m,\omega $. For $\beta_0\sim 10^{50}$ we find the lower
bound
$t_R\geq 10^{-9}s$. However (\ref{tr1}) can be used to restrict $\beta_0$ as
well. Also from (\ref{jj4}) we arrive at a similar bound on $\beta_0$. This is
the independent bound 
on GUP parameter that emerges from our GCS analysis.\\
{\bf{Conclusion and Future Prospects}}: Let us put our work in proper
persective. We emphasize that
until and unless one can (at least theoretically) establish the existence of
stable {\it{quantum states}} 
obeying GUP, the GUP-based phenomenological results \cite{ncr1,ncr} are of
purely academic interest. This Letter provided the first explicit example of
such states in the form of Generalized Coherent States. Hence our work non only
provides this missing link in
GUP-based phenomenology, it also motivates and strengthens earlier results
\cite{ncr1,ncr}.

We have successfully constructed GCS
that satisfies the GUP with its inherent length scale. Various features of these
states
are studied. In subsection {{\bf{1}} of {\bf{Discussion}} we demonstrate the
validity of GUP for the GCS
constructed here and in subsection {{\bf{3}} we derive the independent bound
revealed by our Revival
Time analysis. These results are two of our  most important contributions. 

 A major convenience of our scheme is its computational simplicity:  only
first order corrected energy together with undeformed
Fock basis states are sufficient to generate non-trivial  behaviour of the
perturbed system. This has been clearly shown in the present Letter.

We recall that originally the idea of a GUP and the
associated  minimum
length scale was mooted in \cite{gup1} from high energy string scattering. In
the present
set up we can try to compute scattering cross-sections by using directly the
GCS.
The form of GUP and the NC phase space are precursers
to the relativistic $\kappa$-Minkowski spacetime \cite{nc1}. It will
be interesting if one can construct GCS for the relativistic framework (see
\cite{kai} for the formalism). Lastly our formalism can be exploited in other
physically relevant non-linear oscillator systems \cite{ghosh}.\\
{\bf Appendix:}The results presented in this paper are expressible in closed
form mainly because
we restrict ourselves to $O(\lambda )$ corrections only. Below we provide a
brief outline of the computational scheme.

 From the   Gazeau-Klauder  \cite{gcs} GCS let us
compute $<a>$: 
\begin{equation}
<a>=<J,\gamma \mid a\mid J,\gamma >=\frac{1}{N^2(J)}\sum _{n,m\geq
0}\frac{\sqrt n J^{(n+m)/2}e^{i\gamma
(e_m-e_n)}}{\sqrt
{\rho_m\rho_n}}\delta_{m,n-1}, .
\label{agcs}
\end{equation}
Using $e_{n-1}-e_n=-(1+2n\lambda )$ we find
\begin{equation}
 <a>=\frac{e^{-i\gamma }}{\sqrt J N^2}[\sum _{n}\frac{\sqrt
nJ^{n}e^{-i2n\lambda\gamma
}}{\sqrt
{\rho_{(n-1)}\rho_n}}]$$$$
=\frac{e^{-i\gamma }}{\sqrt J N^2}[\sum _{n}\frac{\sqrt
nJ^{n}e^{-i2n\lambda\gamma
}}{\rho_{(n-1)}\sqrt
{n(1+\lambda (1+n))}}]$$$$
\approx \frac{e^{-i\gamma }}{\sqrt J N^2}[\sum _{n}\frac{J^n(1-i2n\lambda \gamma
)(1-\frac{\lambda}{2}(1+n))}{\rho_{(n-1)}}$$$$
=\frac{e^{-i\gamma }}{\sqrt J N^2}[(1-\frac{\lambda}{2})\sum
_{n}\frac{J^n}{\rho_{(n-1)}}-(\frac{1}{2}+2i\gamma )
\lambda \sum _{n}\frac{nJ^n}{\rho_{(n-1)}}]$$$$
=\sqrt Je^{-i\gamma }[1-\frac{\lambda}{2}\{2+J+4i\gamma (1+J)\}].
\label{a1}
\end{equation}
Calculations for higher order terms in $a,a^\dagger $ follow in a
straightforward manner. We provide another example below.
\begin{equation}
<N>=<a^\dagger a>=\sum_{m,n\geq 0
}\frac{J^{(m+n)/2}e^{i\gamma(e_m-e_n)}}{\sqrt{\rho_m\rho_n}}n\delta_{mn}=\sum_{
n\geq 0}\frac{nJ^n}{\rho_n}$$$$
=\sum_{n\geq 1}\frac{J^n}{\rho_{n-1}(1+\lambda (1+n))}\approx \sum_{n\geq
0}
\frac{J^{n+1}(1-\lambda (1+n))}{\rho_n}=J(1+J)(1-2\lambda (1+J)) .
\label{compu}
\end{equation}\\
{\bf{Acknowledgement}}: One of the authors (S.G) would like to thank Professor
John Klauder for many
helpful correspondences. We thank the referee for constructive comments.

\end{document}